\documentclass[%
 reprint,
bibnotes,
 amsmath,amssymb,
 aps,
]{revtex4-1}

\usepackage[pdftex]{graphicx}
\usepackage{dcolumn}
\usepackage{bm}
\usepackage{theorem}
\theorembodyfont{\normalfont}

\usepackage{color}
\usepackage{comment}
\usepackage{txfonts}
\usepackage{epstopdf}

\newif\iffigure
\figuretrue

\begin{document}

\preprint{APS/123-QED}

\title{Bayesian Spectral Deconvolution Based on Poisson Distribution: Bayesian Measurement and Virtual Measurement Analytics (VMA)}

\author{Kenji Nagata$^{1,2,3}$}
\author{Yoh-ichi Mototake$^4$}
\author{Rei Muraoka$^4$}
\author{Takehiko Sasaki$^4$}
\author{Masato Okada$^{3,4}$}

\affiliation{%
National Institute of Advanced Industrial Science and Technology, Tsukuba, Ibaraki 305-0045, Japan$^1$\\
Japan Science and Technology Agency, PRESTO, Kawaguchi, Saitama 332-0012, Japan$^2$\\
Research and Services Division of Materials Data and Integrated Systems, National Institute for Materials Science, Sengen, Tsukuba, Ibaraki, 305-0047, Japan$^3$\\
Graduate School of Frontier Sciences, The University of Tokyo, 5-1-5 Kashiwanoha, Kashiwa, Chiba 277-8561, Japan$^4$\\
}%




\date{\today}

\begin{abstract}
In this paper, we propose a new method of Bayesian measurement for spectral deconvolution, which regresses spectral data into the sum of unimodal basis function such as Gaussian or Lorentzian functions. 
Bayesian measurement is a framework for considering not only the target physical model but also the measurement model as a probabilistic model, 
and enables us to estimate the parameter of a physical model with its confidence interval through a Bayesian posterior distribution given a measurement data set. 
The measurement with Poisson noise is one of the most effective system
to apply our proposed method. 
Since the measurement time is strongly related to the signal-to-noise ratio for the Poisson noise model, 
Bayesian measurement with Poisson noise model enables us to clarify the relationship between the measurement time and the limit of estimation. 
In this study, we establish the probabilistic model with 
Poisson noise for spectral deconvolution. 
Bayesian measurement enables us to perform virtual and computer simulation for a certain measurement through the established probabilistic model. 
This property is called ``Virtual Measurement Analytics(VMA)" in this paper. 
We also show that the relationship between the measurement time and
the limit of estimation can be extracted by using the proposed method
in a simulation of synthetic data and real data for XPS measurement of MoS$_2$.


\end{abstract}

\pacs{Valid PACS appear here}
\maketitle



\section{Introduction}
Spectral deconvolution is one of the data analysis methods, that regress
the observed spectral data with a multiple-peak structure into the sum of unimodal basis functions
such as Gaussian or Lorentzian functions.
Spectral deconvolution is widely used in several fields of science,
such as physics, chemistry, biology, and planetary science.
In planetary science, the spectral deconvolution is applied 
to the reflectance spectral data of the moon and asteroids to estimate
the mineral composition of their surfaces from the peak positions obtained by the spectral deconvolution. 
In X-ray photoelectron spectroscopy (XPS)\cite{XPSHandBook}, 
the position of each decomposed peak corresponds to the discrete energy level of the inner shell electron 
and is used to extract information reflecting the valence.
By performing spectral deconvolution, we can obtain the physical properties of a certain system 
by extracting the peak parameters such as their position, intensity, and width.
To obtain the peak positions, we need to estimate not only 
peak parameters but also the number of peaks.
Even for the same spectrum data, if the number of peaks used is changed, the estimation result obtained is considerably different,
which leads to a different physical interpretation of the system of interest.

Nagata et al. proposed the Bayesian estimation method for spectral deconvolution\cite{Nagata2012}.
Bayesian estimation is a framework of data analysis, which represents the generation of measurement data as a probabilistic formulation
and estimates the parameters of the generation process by reversing the causality using Bayes' theorem.
They proposed a hierarchical generation model, with which the peak number $K$ 
is stochastically determined and then the parameters of each peak are generated according to $K$.
Nagata et al. used a Gaussian model for spectrum generation.
By applying Bayesian inference to this generation model, they succeeded 
in simultaneously estimating the peak number and peak parameters only from the measurement data.
Moreover, Tokuda et al. extended this framework and succeeded in estimating 
the level of noise added to the spectral data\cite{Tokuda2017}.

In this paper, we propose a new framework of Bayesian spectral deconvolution, 
which treats the generation process of quantum species such as light and photoelectron as a Poisson noise distribution.
In Nagata et al. and Tokuda et al.\cite{Nagata2012,Tokuda2017}, the Bayesian spectral deconvolution approximates the Poisson noise distribution as Gaussian noise distribution.
In this paper, we extend these previous studies so that the numbers of light and photoelectrons are small,
and we deal with the Poisson noise distribution instead of the Gaussian noise distribution.
In order to verify the effectiveness of the proposed method, we also show the results of applying it to measurement data of XPS.
We apply the proposed Bayesian estimation method to several XPS data with different measurement times.
By comparing the results of Bayesian estimation for these data, we show that we can 
estimate the appropriate measurement time to extract the correct latent structure from given spectral data.
This result is valuable because it can be used to acquire an appropriate minimum measurement time in time-resolved XPS\cite{Hagimoto1999,Deng2000}.

The structure of this paper is as follows.
In Sect. 2, we descirbe a framework of Bayesian measurement for spectral deconvolution, and introduce some previous studies of Bayesian spectral deconvolution.
In Sect. 3, we formulate a Bayesian spectral deconvolution that treats the quantum generation process as a Poisson distribution.
In Sect. 4, we perform virtual measurement analytics (VMA) to verify the framework proposed in Sect. 3 using artificial data.
In Sect. 5, we apply the framework proposed in Sect. 3 to molybdenum disulfide (MoS $ _ 2 $).
In Sect. 6, we summarize this paper and describe future developments.

\begin{figure}[b]
\begin{center}
\includegraphics[width=0.9\linewidth]{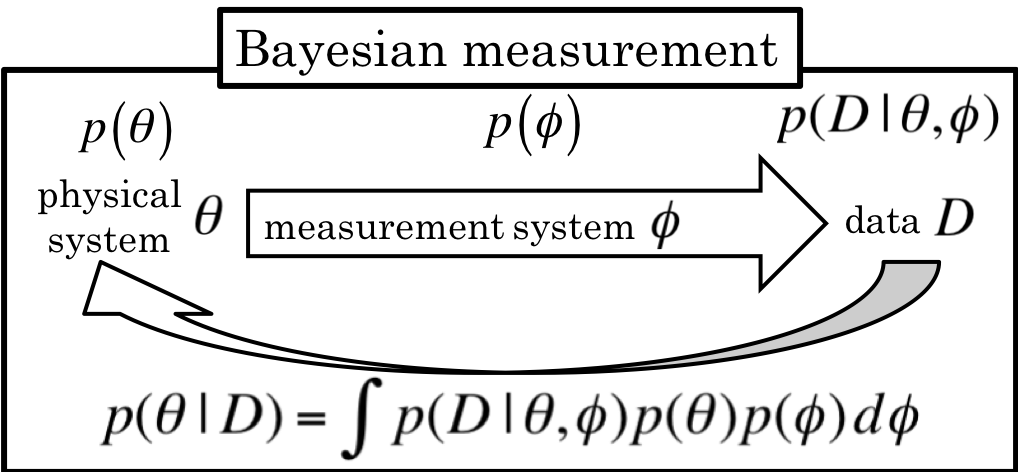}
\caption{Framework of Bayesian measurement.}
\label{BayesianMeasurement}
\end{center}
\end{figure}

\section{Bayesian Measurement for Spectral Deconvolution}
In this section, we describe a framework of Bayesian measurement for spectral deconvolution.
Bayesian measurement is a framework for considering the target physical model $p(\theta)$
and measurement model $p(\phi)$ as probabilistic models.
Bayesian measurement enables us to estimate the physical parameter $\theta$ through 
a Bayesian posterior distribution $p(\theta | D)$ given a measurement data set $D$ with its confidence interval as shown in Fig. \ref{BayesianMeasurement}.

In conventional spectral deconvolution, 
we generate virtual measurement data from a physical model by computer simulation 
and discuss the consistency of the physical model by comparing them with actual measurement data.
If $p(\phi)$ is assumed to be a Gaussian distribution,
then the physical parameter is determined by fitting the experimental and theoretical results by the least-squares method.
That is, $\theta$ is determined by minimizing the squared error $E_m(\theta)$
between the data set $D = \{ x_i, y_i\}$ and the spectral model $f(x;\theta,K)$ given as:
\begin{eqnarray}
E_m(\theta) & = & \frac{1}{2n} \sum_{i=1}^{n} \left[ y_i - f(x;\theta,K) \right]^2.
\end{eqnarray}
If the noise level is too large, 
it is expected to affect the confidence interval of the estimated physical parameter $\theta$.
However, in the conventional spectral deconvolution framework, the confidential interval for the parameter estimation is not considered.
Moreover, in the fitting based on the least squares method, determining the number of peaks objectively is difficult.

Nagata et al. proposed a Bayesian estimation method for spectral deconvolution,
which enables us to estimate not only the physical parameter $\theta$ but also the number of peaks by the model selection method\cite{Nagata2012}.
Bayesian spectral deconvolution can also be used to estimate the confidence interval of the estimated parameter $\theta$
through the posterior distribution $p(\theta | D)$.
Moreover, Tokuda et al. extended Bayesian spectral deconvolution, which enables us to estimate the noise level
as well as the number of peaks\cite{Tokuda2017}.
In these estimation method, noise is assumed to have a Gaussian distribution.
In this study, we propose a framework of Bayesian estimation with Poisson distribution as the measurement model, which is useful for XPS measurement.

\section{Bayesian Spectral Deconvolution with Poisson Distribution}
The spectral function $f(x;\theta,K)$ for the energy value $x \in \mathbb{R}$ is expressed as the sum of 
the signal function $G(x;\theta,K)$ and the background function $B(x;\theta)$,
\begin{eqnarray}
f(x;\theta,K) & := & G(x;\theta,K)+B(x;\theta, K), \label{eq:spectrum} \\
G(x; \theta, K) & := & \sum_{k=1}^{K} a_k \phi(x; \mu_k, \sigma_k).
\end{eqnarray}
The parameter set $\theta$ is $\theta = \{w, v\}, w = \{a_k, \mu_k, \sigma_k\}_{k=1}^{K}$, where
$w$ is the parameter set of the basis function $G(x; \theta, K)$,
and the parameters $a_k, \mu_k, and \sigma_k$ respectively indicate 
the peak intensity, peak position, and width.
The function $\phi(x; \mu_k, \sigma_k)$ represents a unimodal basis function such as 
the Gaussian, Lorentz, or Voigt function.
The definition of the background parameter set $v$ 
depends on the modeling of the background function $B(x; \theta, K)$.
In this study, we focus on the two models of the background function $B(x; \theta, K)$, 
namely, the constant model and the Shirley model\cite{Shirley}.
The constant model is defined as $B(x;\theta, K) = B$, then the parameter set $v$ is $v = \{ B \}$.
In this case, the background function does not depend on the number $K$ of peaks and 
the parameter set $w$ of the signal function.
The Shirley model is another well-known background model, 
which is defined as:
\begin{eqnarray}
B(x; \theta, K) = c \int_{-\infty}^{x} G(u; \theta, K) du + {\rm h}_{start}.
\end{eqnarray}
Then, the parameter set $v$ is $v = \{ {\rm h}_{start}, c\}$, 
where ${\rm h}_{start}$ is the start point of the background 
and the variable $c$ is called background coefficient in this study.

In this study, the generation process is modeled as a Poisson distribution.
Then, the measured spectrum intensity $y \in \mathbb{N}$ for a certain energy $x$ is expressed by 
the following conditional probability distribution $p(y | x, \theta,K)$,
\begin{eqnarray}
p(y | x, \theta, K) & = & \frac{f(x;\theta,K)^{y} \exp\left(-f(x;\theta,K)\right)}{y!}.
\end{eqnarray}
If the data set $D = \{x_i, y_i\}_{i=1}^{n}$ is generated with independent and identically distributed, the probability distribution of the data set $D$ 
is given by 
\begin{eqnarray}
p(D | \theta, K) & = & \prod_{i=1}^{n} p(y_i | x_i, \theta, K) :=  \exp \left( - n E(\theta,K) \right), \\
E(\theta, K) & = & \frac{1}{n} \sum_{i=1}^{n} \left\{f(x_i ; \theta, K) - y_i \log f(x_i ; \theta, K) + \sum_{j=1}^{y_i} j \right\},
\end{eqnarray}
where $n$ is the number of obtained data.

Bayesian spectral deconvolution treats not only the data set $D$ 
but also the parameter set $\theta$ and the number $K$ of peaks
as random variables.
Firstly, the number $K$ of peaks is assumed to be generated subject to the probability $p(K)$.
Next, the parameter set $\theta$ is assumed to be generated with the probability $p(\theta | K)$.
Lastly, the data set $D$ is generated with the conditional probability $p(D | \theta ,K)$.
Then, the joint probability distribution $p(D, \theta, K)$ of all stochastic variables is given by
\begin{eqnarray}
p(D, \theta, K) & = & p(D | \theta, K) p(\theta | K) p(K)
\end{eqnarray}
The posterior probability distribution $p(\theta | D, K)$ of the parameter set $\theta$ 
given the data set $D$ and the number $K$ is expressed using Bayes' theorem as:
\begin{eqnarray}
p(\theta | D, K) & = & \frac{p(D, \theta, K)}{\int p(D, \theta, K) d \theta} \\
 & = & \frac{1}{Z(K)} \exp \left( - n E(\theta,K) \right) p(\theta | K) \label{posterior}, \\
Z(K) & = & \int \exp \left( - n E(\theta,K) \right) p(\theta | K) d \theta.
\end{eqnarray}
The posterior probability distribution $p(K | D)$ is given using Bayes' theorem and the principle of marginalization as:
\begin{eqnarray}
p(K | D) & = & \frac{\int p(D, \theta, K) d \theta}{\sum_{K} \int p(D, \theta, K) d \theta} \\
 & = & \frac{p(K)}{\bar{Z}} \exp \left( - F(K) \right) \label{K_pos},\\
F(K) & = & - \log \int \exp \left( - n E(\theta,K) \right) p(\theta | K) d \theta, \\
\bar{Z} & = & \sum_{K} \int \exp \left( - n E(\theta,K) \right) p(\theta | K) p(K) d \theta,
\end{eqnarray}
where $\sum_{K}$ indicates the sum for all possible number $K$ of peaks.
Consequently, Bayesian estimation calculates the posterior distribution through 
the log loss function $E(\theta, K)$ and the free energy $F(K)$.
In this study, we consider the estimation method, 
which firstly estimates the number $K$ of peaks by maximizing the posterior distribution $p(K|D)$,
and which next estimates the parameter set $\theta$ by maximizing the posterior distribution $p(\theta|K,D)$.

\begin{figure*}[t]
\begin{center}
\includegraphics[width=1.0\linewidth]{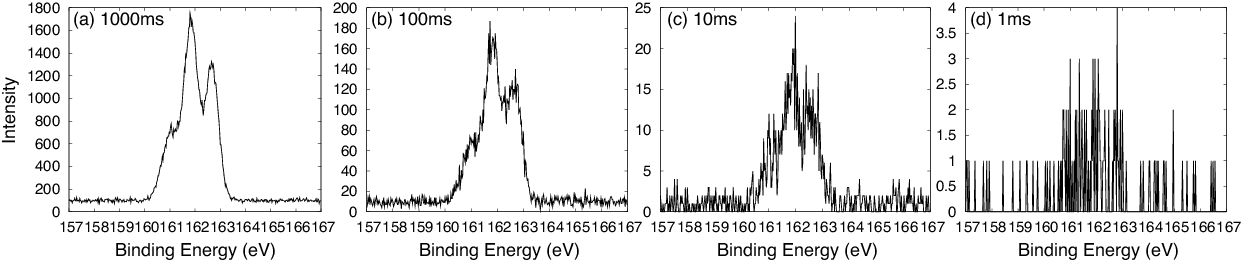}
\caption{Examples of the artificial spectral data in this study.}
\label{graph:XPS}
\end{center}
\end{figure*}

We use the exchange Monte Carlo (EMC) method to simulate Bayesian estimation\cite{Nagata2012,Hukushima1996}.
This method is one of the Markov chain Monte Carlo (MCMC) methods.
Let us consider the sampling from the posterior distribution $p(\theta | D, K)$ shown in Eq. (\ref{posterior}).
In the EMC method, we prepare the replicated probability distribution $p_{\beta}(\theta | D, K)$ with the inverse temperature $\beta$ as:
\begin{eqnarray}
p_{\beta}(\theta | D, K) & \propto & \exp \left( -n \beta E_p(\theta, K) \right) p(\theta | K). \label{replica}
\end{eqnarray}
Hence, the target distribution for the EMC method is the following joint distribution with the different inverse temperatures $0 \leq \beta_1 \leq \cdots \leq \beta_M \leq 1$:
\begin{eqnarray}
p(\theta_1, \ldots, \theta_M) & = & \prod_{m=1}^M p_{\beta_m}(\theta_m | D, K).
\end{eqnarray}
The algorithm of the EMC method consists of the following two updates: 
\begin{description}
\item [Step 1: Update for each replica]\mbox{}\\
Update the parameter set $\theta$ for each replicated probability distribution $p_{\beta_m}(\theta | D, K)$ using the Metropolis algorithm,
which is the most fundamental algorithm of the MCMC method.
\item[Step 2: Exchange between the neighboring replicas]\mbox{}\\
Exchange the current states of the parameter set $\theta$ for the neighboring replicas,
that is, $\{ \theta_m, \theta_{m+1} \} \to  \{ \theta_{m+1}, \theta_{m}  \}$, 
with the following probability $p(\theta_m\leftrightarrow \theta_{m+1})$:
\begin{align}
&p(\theta_m\leftrightarrow \theta_{m+1}) = \min(1, v), \\
&v= \frac{p_{\beta_m}(\theta_{m+1} | D, K)p_{\beta_{m+1}}(\theta_m | D, K)}{p_{\beta_m}(\theta_m | D, K)p_{\beta_{m+1}}(\theta_{m+1} | D, K)} \\
&\phantom{v}= \exp\left( (\beta_{m+1}-\beta_m)(E_p(\theta_{m+1},K)-E_p(\theta_m,K)) \right).
\end{align}
\end{description}
By the update procedure, we obtain the sample sequence from the joint distribution shown in Eq. (\ref{replica}).
We obtain the estimated parameter $\theta$ by maximizing the posterior distirbution $p(\theta | D, K)$ 
from the obtained sample sequence.
The EMC method enables us to sample the global optimal solution efficiently
without trapping at local minimum solutions owing to the exchange process in the EMC method.

We can also calculate the partition function in Eq. (\ref{K_pos}) efficiently\cite{Nagata2012, Ogata1990}.
Let us consider the following function $z(\beta)$ defined as
\begin{eqnarray}
z(\beta) & = & \int d\theta \exp \left( - n \beta E_p(\theta, K) \right) p(\theta | K).
\end{eqnarray}
From the definition, it is clear that $z(0) = 1$.
The aim is to calculate the value of the function $z(1)$, which corresponds to the partition function.
This value is given by using a sequence of inverse temperatures $0 \leq \beta_1 \leq \cdots \leq \beta_M \leq 1$ as follows:
\begin{eqnarray}
z(1) & = & \frac{z(\beta_M)}{z(\beta_{M-1})} \times \cdots \times \frac{z(\beta_2)}{z(\beta_1)} = \prod_{m=1}^{M-1} \frac{z(\beta_{m+1})}{z(\beta_m)} \nonumber \\
 & = & \prod_{m=1}^{M-1} \frac{\int d\theta \exp \left( - n \beta_{m+1} E_p(\theta, K) \right) p(\theta | K)}{\int d\theta \exp \left( - n \beta_m E_p(\theta, K) \right) p(\theta | K)} \nonumber \\
 & = & \prod_{m=1}^{M-1} \left< \exp \left( - n (\beta_{m+1} - \beta_{m}) E_p(\theta, K) \right)\right>_{p_{\beta}(\theta | D, K)},
\end{eqnarray}
where the notation $\left< \cdot \right>_{p_{\beta}(\theta | D, K)}$ shows the expectation over the probability distribution $p_{\beta}(\theta | D, K)$.
These expectations can be calculated by using the sample sequence generated by the EMC method.
Consequently, the EMC method enables us to not only search for the optimal parameter set $\theta$ 
but also calculate the free energy $F(K)$ as $F(K) = - \log z(1)$, 
which is important for the choice of the number $K$ of peaks.

\section{Validation of our Proposed Method by VMA}
As mentioned earlier, for the Bayesian measurement for spectral deconvolution,
we can virtually generate measurement data from the target physical model $f(x;\theta,K)$
and the measurement model $p(y|x,\theta,K)$ by computer simulation.
Virtual measurement analytics (VMA) is an information mathematical concept that
performs virtual and computational simulations in order to generate measurement data from
the assumed physical and measurement models.
Using VMA, we can check the confidence interval of the physical parameter $\theta$ through the posterior distribution.
In, in this section, we demonstrate the result of VMA for Bayesian spectral deconvolution 
defined earlier.

For the generated data, the basis function $\phi(x;\mu,\sigma)$ of the signal function $G(x; \theta)$ 
was set as a Gaussian function,
\begin{eqnarray}
\phi(x;\mu,\sigma) & = & \exp \left( \frac{(x-\mu)^2}{2 \sigma^2}\right) \label{Gauss}.
\end{eqnarray}
The background $B(x;\theta, K)$ was set as a constant function, $B(x;\theta, K) = B$.
The number of peaks was set as 3, 
and the true parameter $\theta^* = \left\{\{a_k^*,\mu_k^*,\sigma_k^*\}_{k=1}^{3}, B^*\right\}$ was set as:
\begin{eqnarray}
\left( 
\begin{array}{c}
a_1^*
\\
a_2^*
\\
a_3^*
\end{array}
\right) =
T \left(
\begin{array}{c}
0.587
\\
1.522
\\
1.183
\end{array}
\right) & , &
\left(
\begin{array}{c}
\mu_1^*
\\
\mu_2^*
\\
\mu_3^*
\end{array}
\right) = 
\left(
\begin{array}{c}
161.032
\\
161.851
\\
162.677
\end{array}
\right) \nonumber \\
\left( 
\begin{array}{c}
\sigma_1^*
\\
\sigma_2^*
\\
\sigma_3^*
\end{array}
\right) = 
\left(
\begin{array}{c}
0.341
\\
0.275
\\
0.260
\end{array}
\right) & , & 
B^* = 0.1 \times T \nonumber
\end{eqnarray}
where $T$ is a pseudo-measurement time.
The longer the measurement time, the higher the obtained intensity,
which leads to a good signal-to-noise ratio.
The pseudo-measurement time was set as four patterns, $T = \{1.0, 10.0, 100.0, 1000.0\}$.
Figure \ref{graph:XPS} shows examples of the generated spectral data for (a) $T=1000.0$, (b) $T=100.0$, (c) $T=10.0$, and (d) $T=1.0$ (ms).
The spectral data in Fig. \ref{graph:XPS}(a) can be observed to have three peaks owing to the good signal-to-noise ratio.
In constarst, the spectral data in Fig. \ref{graph:XPS}(d) have a high level of noise, and it is difficult to judge if these this data have three peaks.

In this study, we generated 50 patterns of spectral data for each measurement time.
We estimated the number $K$ of peaks on the basis of the minimization of the free energy for each data in order to clarify the minimum measurement time to determine the correct number $K$ of peaks.
The candidate number of peaks was set from $K=1$ to $5$.
That is, the prior distribution $p(K)$ was set as the discrete uniform distribution from $K=1$ to $5$.
The prior distribution $p(\theta|K)$ of the parameter set $\theta$ was set by using the gamma and Gaussian distributions 
on the basis of the domain and the property of each parameter as:
\begin{eqnarray}
p(\theta|K) & = & \prod_{k=1}^{K} p(a_k) p(\mu_k) p(\sigma_k) p(B), \\
p(a_k) & = & {\rm Gamma}(a_k; \eta_{a}, \lambda_{a}) \\
 & = & \frac{1}{\Gamma(\eta_{a})}(\lambda_{a})^{\eta_{a}} (a_{k})^{\eta_{a}-1} \exp(-\lambda_{a}a_{k}), \\
p(\mu_k) & = & N\left(\mu_{k};\nu_{0},\xi_{0}^2\right) \\
 & = & \frac{1}{\sqrt[]{2\pi \xi_{0}^2}}\exp \left(\frac{(\mu_{k}-\nu_{0})^2}{2\xi_{0}^2}\right), \\
p(\sigma_k) & = & {\rm Gamma} \left( \frac{1}{\sigma_k^2} ; \eta_{\sigma}, \lambda_{\sigma}\right), \\
p(B) & = & N\left(B; \nu_{B}, \xi_{B}^2\right). 
\end{eqnarray}
The parameters for the prior distribution $p(\theta |K)$ were set as:
\begin{eqnarray}
\eta_a = 2.0 & , & \lambda_a = 2.0 \times T, \\
\nu_0 = 160.0 & , & \xi_0 = 2.0, \\
\eta_{\sigma} = 10.0 & , & \lambda_{\sigma} = 2.5, \\
\nu_{B} = 0.1 \times T & , & \xi_{B} = 0.01 \times T
\end{eqnarray}
For the setting of the EMC method, the number $M$ of replicas was set as 32,
and the inverse temperature $\beta_m$ of each replica was set as:
\begin{equation}
 \beta_m = 
 \left\{
  \begin{aligned}
   &0 \; (m = 1)\\
   &1.5^{m-M} \; (m\geq 2).
  \end{aligned}
 \right.
\end{equation}
This setting is based on the theoretical knowledge 
that the exponential setting of inverse temperatures makes the acceptance rate of the exchange process constant for inverse temperatures at a low temperature limit\cite{Nagata2007}.

\begin{table}[t]
\begin{center}
\begin{tabular}{|c|c|c|c|c|c|}
\hline
Measurement time (ms) & $K=1$ & $K=2$ & $K=3$ & $K=4$ & $K=5$\\
\hline
(a)1000 & 0 & 0 & $\bf{50}$ & 0 & 0 \\
\hline
(b)100 & 0 & 0 & $\bf{50}$ & 0 & 0 \\
\hline
(c)10 & 0 & 3 & $\bf{47}$ & 0 & 0 \\
\hline
(d)1 & 1 & 28 & $\bf{20}$ & 1 & 0 \\
\hline
\end{tabular}
\caption{Frequency of model selection based on the free energy $F(K)$ for each measurement time.
The numbers in bold show the frequency of choosing the correct number $K$ of peaks.}
\label{model_selection}
\end{center}
\end{table}

Table \ref{model_selection} shows the result of the estimation of the number $K$ of peaks for each setting of measurement time.
The numbers in bold show the frequency of choosing the correct number $K$ of peaks.
When the number in bold is large, the corresponding measurement time is suitable for estimating the correct number $K$ of peaks.
From the result in Table \ref{model_selection}, we can see that the Bayesian estimation can be used to estimate the correct number $K$ of peaks
for the spectral data with the measurement time $T=1000.0$, $100.0$, and $10.0$.
In contrast, for $T=1.0$ the estimated number $K$ of peaks varies, 
which indicates that the spectral data with the measurement time $T=1.0$ were insufficient for estimating the correct number $K$ of peaks.

\begin{figure*}[t]
\begin{center}
\includegraphics[width=1.0\linewidth]{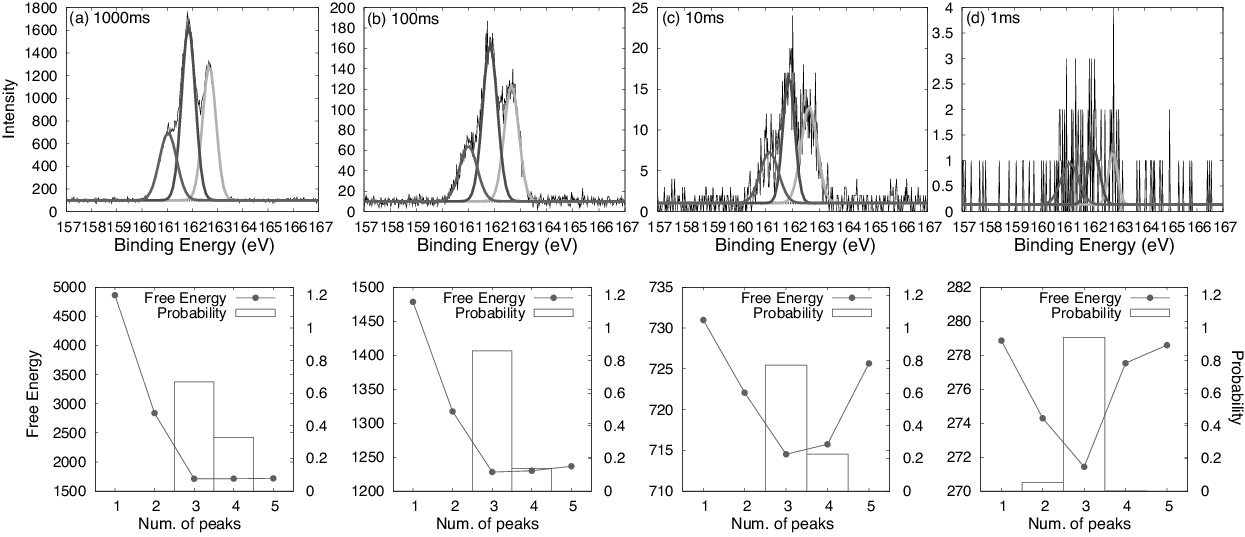}
\caption{Result of Bayesian estimation for the spectral data shown in Fig. \ref{graph:XPS}.
The upper figures show the fitting results for the estimated number $K$ of peaks,
and the lower ones show the free energy $F(K)$ for each number $K$ of peaks.
In the lower figures, the solid lines indicate the free energy $F(K)$, and the bar graphs indicate 
the posterior probability distribution $p(K|D)$ calculated using Eq. (\ref{K_pos}).
In these cases, the optimal number of peaks based on the Bayesian estimation is three in all figures.}
\label{graph:XPSresult}
\end{center}
\end{figure*}

\begin{figure*}[t]
\begin{center}
\includegraphics[width=1.0\linewidth]{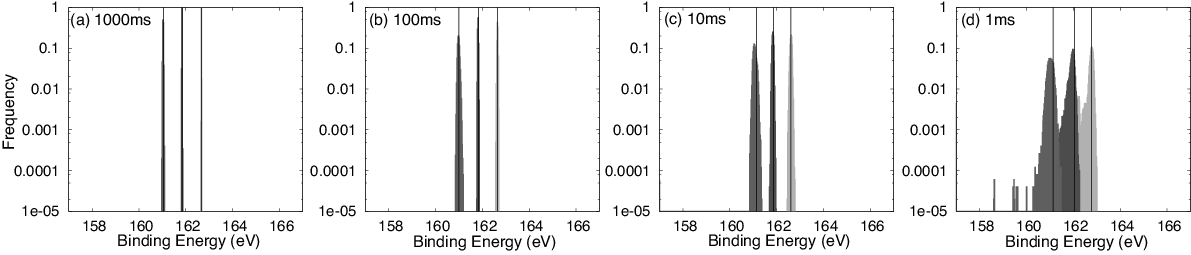}
\caption{Histograms of the posterior probability distribution $p(\mu_k | D, K)$ of the three peak positions $\mu_1$, $\mu_2$, and $\mu_3$, 
for the spectral data shown in Fig. \ref{graph:XPSresult}.}
\label{graph:histogram}
\end{center}
\end{figure*}

Next, we show the detailed result of the Bayesian estimation for each spectral data set shown in Fig. \ref{graph:XPS}, which is shown in Fig. \ref{graph:XPSresult}.
The upper figures show the fitting results for the estimated number $K$ of peaks,
and the lower ones show the free energy $F(K)$ each the number $K$ of peaks.
In the lower figures, the solid lines indicate the free energy $F(K)$, and the bar graphs indicate 
the posterior probability distribution $p(K|D)$ calculated using Eq. (\ref{K_pos}).
As seen from this result, the Bayesian estimation can be used to choose the correct number $K$ of peaks, $K=3$,
and the peak positions $\mu_k$ are also correctly estimated.

Then, we focus on the posterior distribution of the peak position $\mu_k$ in order to validate the confidence of the estimation.
Since Bayesian spectral deconvolution simulates the sampling from the posterior probability distribution $p(\theta | K)$ of the parameter set $\theta$ by the EMC method,
we can evaluate the confidence interval of the parameter $\theta$ by constructing the histogram of the parameter $\theta$ from the sampling result.
Figure \ref{graph:histogram} shows the histograms of the peak positions $\mu_k$ described in Fig. \ref{graph:XPSresult}.
From this figure, the histograms of the three peak positions in Fig. \ref{graph:histogram}(d) overlap each other,
which indicates the inaccuracy of the parameter estimation.
From this result, we can obtain not only the estimation value of the parameter but also its confidence interval 
by checking the posterior distribution.

\section{Validation of Real XPS Data for MoS$_2$}
\begin{figure*}[t]
\begin{center}
\includegraphics[width=0.8\linewidth]{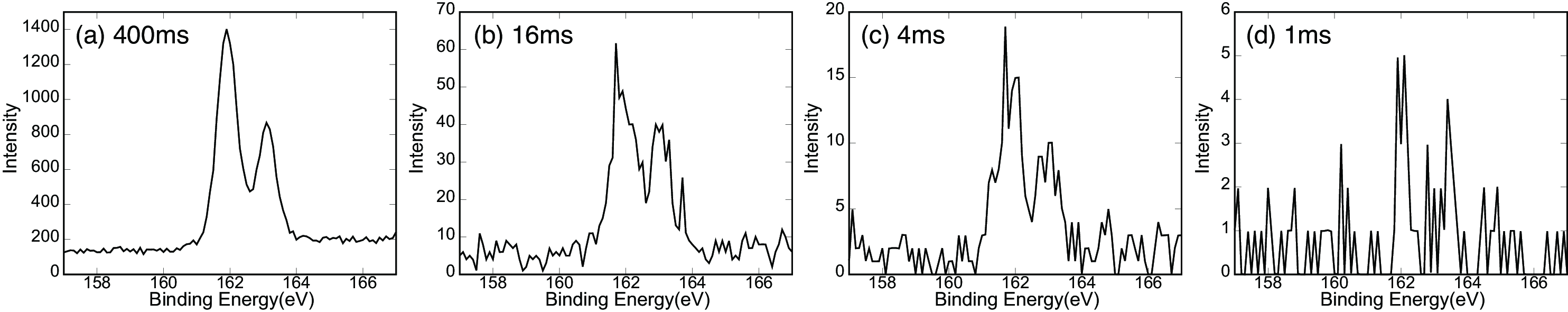}
\caption{Spectral data of MoS$_2$ powder. The measurement times for one step are  
(a) 400, (b) 16, (c) 4 and (d) 1ms.}
\label{graph:MoS2}
\end{center}
\end{figure*}

Next, we show the result of the Bayesian estimation for real XPS data for MoS$_2$,
which is well known as a standard sample for XPS measurement.
Figure \ref{graph:MoS2} shows the measured spectral data of MoS$_2$.
In our XPS measurement, we varied the measurement time and investigated how the noise level affects the estimation accuracy.
The measurement time for one step in each figure was set as 400, 16, 4, and 1 ms.
The sample used in the measurement was molybdenum disulfide (MoS$_2$) powder fixed with carbon tape.
We used a PHI 5000 VersaProbe microprobe (ULVAC-PHI .inc) for the XPS measurement.
The energy range for the measurement was set from 157 to 167 eV, in which the S2p peak of MoS$_2$ exists.
The pass energy was set as 23.50 eV.
The X-ray source used in the XPS measurement was ${\rm AlK_{\alpha}}$ and impurities were removed by a monochromator.
As seen from Fig. \ref{graph:MoS2}(a), this spectrum has two peaks,
which are the peaks of $S2p_{1/2}$ and $S2p_{3/2}$\cite{Baker1999}.

The basis function $\phi(x;\mu,\sigma)$ of the signal function $G(x;\theta, K)$ was used in the following function instead of the Gaussian function in Eq. (\ref{Gauss}) as follows:
\begin{eqnarray}
\phi(x; \mu, b) & = & \frac{\exp \left( -0.3 \times \ln 2 \times b (x - \mu)^2 \right)}{1 + 0.7 \times b (x - \mu_k)^2}.
\end{eqnarray}
This function is a mixed function of the Gaussian and Lorentz functions with a 70:30 ratio 
and is often used for the peak fitting of XPS spectral data\cite{Wahab2007}.
The parameters of the basis function are the peak position $\mu$ and the inverse of the peak variance, $b_k$.
The background $B(x;\theta, K)$ was used in the Shirley model\cite{Shirley}.
The Shirley model is based on the concept that the background arises from photoelectrons
that are ejected by high-energy bremsstrahlung and undergo subsequent energy loss in the sample.
Hence, the background $B(x;\theta,K)$ is given by the following function:
\begin{eqnarray}
B(x;\theta,K) = c \int_{-\infty}^{x} G(u;\theta,K) du + {\rm h}_{start}.
\end{eqnarray}
The parameter set $v$ of the background $B(x;\theta,K)$ is the start point ${\rm h}_{start}$ of the background 
and the variable $c$, which is called background coefficient in this study.
In the typical Shirley model, the variable $c$ is determined by the intensity difference
between the start point and the end point of the background.
In contrast, in this study, we estimate the variable $c$ directly.
Hence, the parameter set $\theta$ is $\theta = \left\{  \{a_k, \mu_k, \sigma_k \}_{k=1}^{K}, c, {\rm h}_{start}\right\}$.
The prior distribution $p(\theta|K)$ of the parameter set $\theta$ was set by using the gamma and Gaussian distributions 
on the basis of the domain and the property of each parameter as:
\begin{eqnarray}
p(\theta|K) & = & p(c) p({\rm h}_{start}) \prod_{k=1}^{K} p(a_k) p(\mu_k) p(\sigma_k), \\
p(a_k) & = & {\rm Gamma}(a_k; \eta_{a}, \lambda_{a}), \\
p(\mu_k) & = & N\left(\mu_{k};\nu_{0},\xi_{0}\right), \\
p(\sigma_k) & = & {\rm Gamma} \left( \frac{1}{\sigma_k^2} ; \eta_{\sigma}, \lambda_{\sigma}\right), \\
p(c) & = & {\rm Gamma}\left(c; \eta_{c}, \lambda_{c}\right), \\
p({\rm h}_{start}) & = & N \left( {\rm h}_{start}; \nu_{start}, \xi_{start} \right).
\end{eqnarray}
The parameters for the prior distribution $p(\theta |K)$ were set as:
\begin{eqnarray}
\eta_a = 2.0 & , & \lambda_a = 2.0 \times T, \\
\nu_0 = 160.0 & , & \xi_0 = 5.0, \\
\eta_{\sigma} = 10.0 & , & \lambda_{\sigma} = 0.4, \\
\eta_c = 0.8 & , & \lambda_c = 0.8, \\
\nu_{start} = 0.35 \times T & , & \xi_{start} = 0.1 \times T.
\end{eqnarray}
For the setting of the EMC method, the number $M$ of replicas was set as 64,
and the inverse temperature $\beta_m$ of each replica was set as:
\begin{equation}
 \beta_m = 
 \left\{
  \begin{aligned}
   &0 \; (m = 1)\\
   &1.25^{m-M} \; (m\geq 2).
  \end{aligned}
 \right.
\end{equation}

\begin{figure*}[t]
\begin{center}
\includegraphics[width=0.9\linewidth]{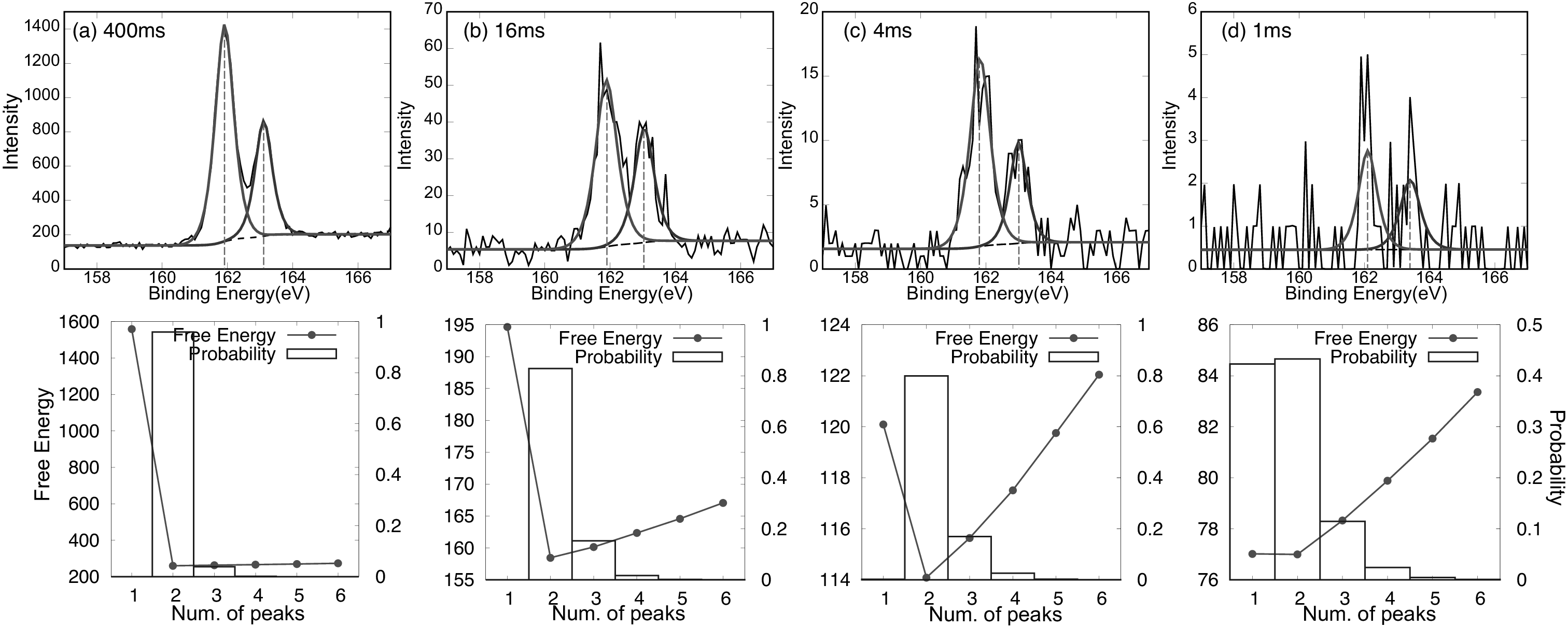}
\caption{Result of Bayesian estimation for the spectral data shown in Fig. \ref{graph:MoS2}.
The upper figures show the fitting results for the estimated number $K$ of peaks,
and the lower ones show the free energy $F(K)$ for the number $K$ of peaks.
In the lower figures, the solid lines indicate the free energy $F(K)$, 
and the bar graphs indicate the posterior probability distribution $p(K|D)$ calculated using Eq. (\ref{K_pos}).}
\label{graph:fitting}
\end{center}

\begin{center}
\includegraphics[width=0.9\linewidth]{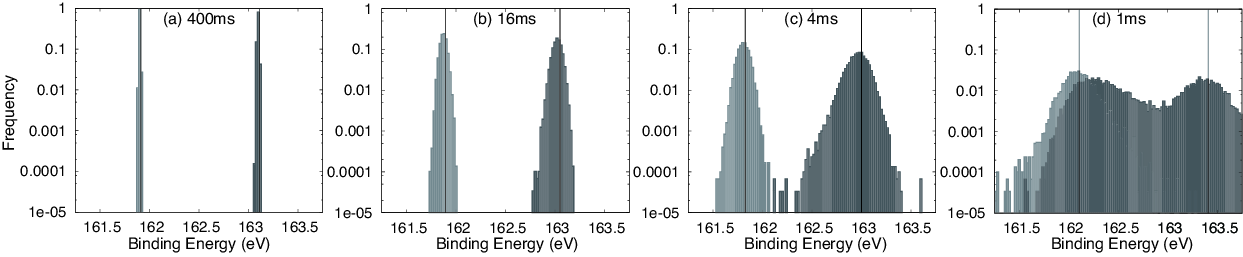}
\caption{Histograms of the posterior probability distribution $p(\mu_k|D,K)$ of the two peak positions $\mu_1$ and $\mu_2$ 
for the spectral data shown in Fig. \ref{graph:fitting}.}
\label{graph:posterior}
\end{center}
\end{figure*}

Figure \ref{graph:fitting} shows the result of the Bayesian estimation for the spectral data shown in Fig. \ref{graph:MoS2}.
The upper figures in Fig. \ref{graph:fitting} show the fitting result with the estimated number $K$ of peaks,
and the lower ones show the result of the free energy $F(K)$.
In the lower figures, the solid lines indicate the value of the free energy $F(K)$, and the bar graphs indicate the posterior probability distribution $p(K|D)$
calculated using Eq. (\ref{K_pos}).
From the result of the free energy $F(K)$, the Bayesian estimation chooses an appropriate number $K$ of peaks, $K=2$, for any spectral data.
According to these results, the Bayesian estimation seems to obtain the correct peak structure,
such as the number $K$ of peaks and the parameters.

\begin{figure*}[t]
\begin{center}
\includegraphics[width=0.6\linewidth]{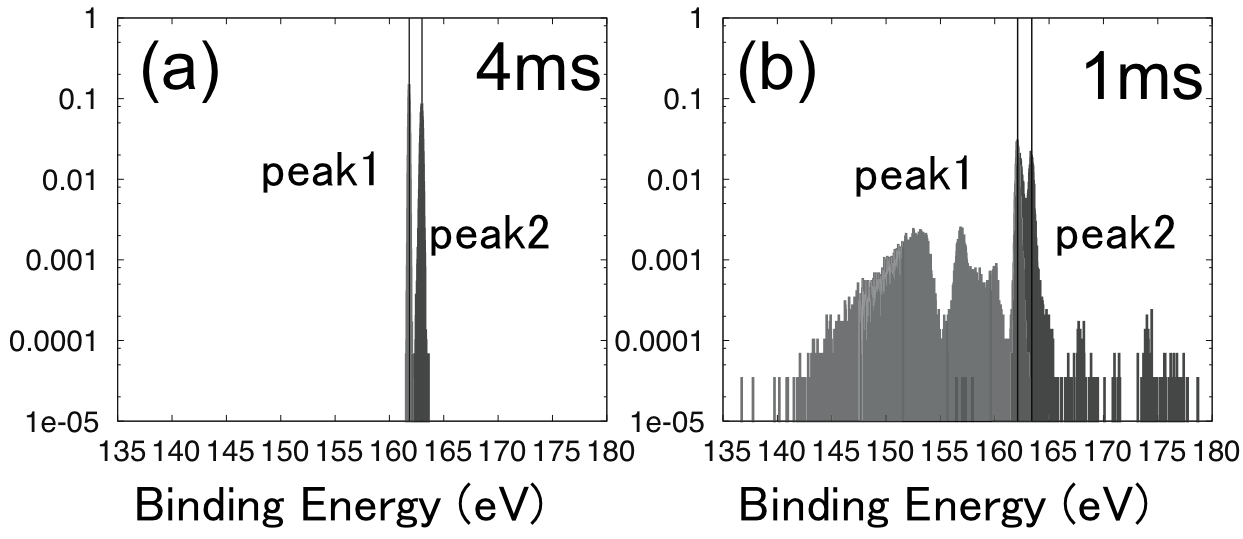}
\caption{Histograms of the posterior probability distribution $p(\mu_k|D,K)$ shown in Figs. \ref{graph:posterior}(c) and \ref{graph:posterior}(d) with a wide range of binding energy.}
\label{graph:posterior02}
\end{center}
\end{figure*}

Figure \ref{graph:posterior} shows the histograms of the two peak positions $\mu_1$ and $\mu_2$ for the fitting shown in Fig. \ref{graph:fitting}.
For the spectral data with a high signal-to-noise ratio such as those in Figs. \ref{graph:MoS2}(a) and \ref{graph:MoS2}(b),
the obtained histograms are narrow, which show the high accuracy of parameter estimation.
In contrast, the histograms for the spectral data in Fig. \ref{graph:MoS2}(d) with a 1 ms measurement time overlap with each other, 
which show the inaccuracy of parameter estimation.
From this result, the spectral data with 400, 16, and 4 ms measurement times are sufficient to be appropriately extracted into two peaks of $S2p_{1/2}$ and $S2p_{3/2}$.
Figure \ref{graph:posterior02} shows the histograms shown in Figs.\ref{graph:posterior}(c) and \ref{graph:posterior}(d) with a wide binding energy (eV) range.
From the result, the histogram for 1 ms is widely spread beyond the energy range for this measurement from 157 to 167 eV.
This indicates that there is a large difference in the accuracy of the parameter between 4 and 1 ms.

Consequently, the Bayesian estimation enables us to clarify the minimum measurement time to appropriately extract the peak structure from XPS data.
This is very useful for designing the XPS measurement.

\section{Summary}

In this study, we proposed a framework of Bayesian measurement, in which we consider not only 
the target physical model but also the measurement model as a probabilistic model.
We also apply the Bayesian measurement to spectral deconvolution, in which we regress 
the spectral data into the sum of basis functions such as the Gaussian function.
By assuming Poisson noise for the measurement noise model such as XPS,
we can estimate not only the model parameter but also noise intensity\cite{Nagata2012,Tokuda2017}.
In the model considering the Poisson noise, the measurement time is strongly related to the signal-to-noise ratio.
Then, we performed Bayesian spectral deconvolution of some spectral data sets with different measurement times as synthetic data and real XPS data for MoS$_2$,
and we confirmed that the relationship between the measurement time and
the limit of estimation can be extracted by using the proposed method.


\section*{Acknowledgment}
This study was partially supported by JSPS Grants-in-Aid for Scientific Research (Grant Numbers 20120009 and 25330293).

\end{document}